\documentstyle[epsf,osa]{revtex}
\begin{document}
\hsize 6.0 truein

\twocolumn[

\title{ Localization in Strongly Chaotic Systems }
\author{\large Mario Feingold}
\address{ Dept. of Physics, Ben-Gurion University, Beer-Sheva
84105, Israel}
\maketitle 
\begin{abstract}
{\widetext We show that, in the semiclassical limit and whenever the 
elements of the Hamiltonian matrix are random enough, the
eigenvectors of strongly chaotic time-independent systems in 
ordered bases can on average be exponentially localized across the
energy shell and decay faster than exponentially outside the energy shell.
Typically however, matrix elements are strongly correlated leading
to deviations from such behavior.}
\end{abstract}
\medskip

\bigskip

\leftline{submitted to {\sl Phys. Rev. Lett., April 1996}} 
\vskip 8pt
\vskip 60pt
]

\hsize 7.0 truein
\narrowtext

The quantum mechanical behavior of strongly chaotic systems 
is commonly thought to be of two, apparently different types.
On one hand, there are the time-periodic systems, e.g. the Kicked Rotor, 
for which the eigenstates of the one period evolution operator, $U$, are
under certain conditions localized.\cite{fish} When this occurs, there is
no repulsion between 
most levels and the spectrum of quasienergies is characterized by a
Poisson spacings distribution.\cite{f4} One the other hand, one has the
time-independent systems, e.g. the Coupled Quartic Oscillators, where
eingenstates in a phase space representation, are on average homogeneously
spread over the corresponding energy shell. Such states are being
regarded as extended leading to strong level repulsion and a Wigner
spacing distribution.\cite{bohi} It is the purpose of this Letter to show
that as far as the behavior of the eigenstates and in particular their
localization properties, the two types of systems are in fact quite
similar and that localization can occur under certain conditions in
time-independent systems as well.

For simplicity, in what follows we shall contain our discussion to a
subclass of the time-periodic systems, namely, to kicked systems 
of the type
\begin{equation}
H = K(p) + V(q) \sum_{n= -\infty}^{\infty} \delta (t - nT) \ .
\end{equation}
Such systems can be mapped onto Anderson models for the motion of
electrons on a purely one-dimensional disordered lattice. Whenever the
corresponding hopping elements, $W_n$, decay faster than $n^{-1}$ and the 
the site energies, $\epsilon_n$, are random enough, the resulting
eigenstates of the Anderson model are known to be exponentially localized.
A more direct approach is based on the fact that in the basis of eigenstates
of $K(p)$ the evolution operator, $U_{nm}$, is a banded matrix with elements
that have pseudo-random phases. Therefore, its eigenvectors are expected
to behave similarly to those of a banded random matrix\cite{cas1} of the same
band width, $b$. Since the latter can be analyzed using a transfer matrix 
formalism, the eigenvectors are exponentially localized with a 
localization length, $\xi$, that is proportional to $b^2$, $\xi = \gamma
b^2$.

The Kicked Rotor for example, corresponding to $K(p) = p^2/2$ and $V(q) =k
\cos q$, was originally introduced as a simplified but representative Poincare 
map of a time-independent system with two degrees of freedom, $d=2$, also 
known as the Standard Map. It is therefore natural to expect that also its
quantum mechanics should be reminiscent of that of time-independent 
systems. One obstacle that has prevented the search for this similarity is 
the fact that eigenstates of time-independent systems are best understood in
phase space representations through the Berry-Voros conjecture.\cite{ber1}
However,
in order to obtain an analogous description to that leading to the
banded evolution operator, $U_{nm}$, a basis with a natural ordering is
required. Let, $H = H_0 +V$, be an
arbitrary separation of $H$. Using the eingenvectors of $H_0$, $v_n$,
arranged in increasing order of the corresponding eigenvalues, $E_{0,n}$,
one obtains for $H$ a matrix representation, $H_{nm}$, that is banded and
moreover, has diagonal elements which vary on classical energy scales.
Although the latter feature which is related to energy conservation is
absent in the $U_{nm}$ matrix where diagonal elements are on average
of unit absolute value, we find that in the semiclassical limit,
$\hbar \to 0$, the
influence of the diagonal on the behavior of the eigenvectors is partially
suppressed. As a consequence, in such bases, the semiclassical global
structure
of the $U_{nm}$ and $H_{nm}$ matrices is very much alike. On the other hand,
a simple mechanism that generates randomness in the $U_{nm}$ matrix is 
absent in the case of $H_{nm}$, making the latter significantly less random.

We now turn to study the structure of the $H_{nm}$ matrix. Let us assume 
that all the relevant operators, namely, $H$, $H_0$ and $V$, are well behaved 
functions of the canonical variables, ${\bf z} \equiv ({\bf q}, {\bf p} ) $.
Then, under quite general conditions, it is known\cite{diag}
that the energy average of the diagonal elements equals to the 
corresponding microcanonical average, that is,
\begin{equation}
<H_{nn}> \simeq \{H ({\bf z})\}_{E_{0,n}}  \ , \label{eq:diag}
\end{equation}
where for any function, $F ({\bf z})$,
\begin{equation}
\{ F ({\bf z}) \}_{E_0} \equiv {{ \int d {\bf z} F ({\bf z}) \delta [E_0 -
H_0({\bf z})] }\over { \int d {\bf z} \delta [E_0 - H_0({\bf z}])}} \ .
\end{equation}
Moreover, the variance of the diagonal elements,
${\nobreak \sigma_D^2 \equiv <H_{nn}^2> -<H_{nn}>^2 }$,
was shown\cite{f8} to be equal to twice the
variance of the off-diagonal matrix elements that are close to the diagonal,
$\sigma_O^2$, and
correspondingly, $\sigma_D^2 = O(\hbar^{d-1})$. Therefore, in the
semiclassical limit the distribution of diagonal elements has vanishing 
width whenever $d >1$. As we now show, this fact is extremely helpful
in understanding the behavior of the off-diagonal matrix elements of $H_{nm}$.
In the $n$-th row of $H_{nm}$, the average distance of matrix elements from
the diagonal measured in units of energy is
\begin{eqnarray}
(\Delta E_{0,n})^2 &\equiv& {{\sum_m \ (E_{0,m} - E_{0,n})^2 | H_{nm} |^2}
\over{\sum_{m(\not= n)} | H_{nm} |^2}} \ \nonumber \\
& = & {{([H_0, H]^2)_{nn}}\over
{(H^2)_{nn} - (H)_{nn}^2}} \ , \label{eq:del}
\end{eqnarray}
and using Eq. (\ref{eq:diag})
\begin{equation}
<(\Delta E_{0,n})^2> \ \to \ \hbar^2 { { \{ [H_0, H]_{PB}^2 \} }\over
{ \{ H^2 \}  - \{ H \} ^2 }}  \quad {\rm for } \ \hbar \to 0 \ ,
\label{eq:dels}
\end{equation}
where the commutator of Eq. (\ref{eq:del}) was replaced by $\hbar$ times
the corresponding Poisson bracket, $[...]_{PB}$. Notice that while Eq. 
(\ref{eq:diag}) holds for arbitrary values of $\hbar$,
Eq. (\ref{eq:dels}) only applies for small enough $\hbar$. The reason is
that in Eq. (\ref{eq:dels}), the lhs is the average of a ratio 
of diagonal elements which, in general, differs from the ratio of
the averages on the rhs. However, if $d>1$ and $\hbar \to 0$,
then the width of 
the distribution of the diagonal elements appearing in the denominator
of Eq. (\ref{eq:del}) becomes vanishingly small, therefore playing the
role of a constant in the averaging.

In order to fully grasp the implication of Eq. (\ref{eq:dels}), it is 
necessary to express $<(\Delta E_{0,n})^2>^{1/2}$ in terms of the number of
states by multiplying it with the mean density of states, $\rho(E_0)$,
which is $O(\hbar^{-d})$, such that $<(\Delta N_{0,n})^2>^{1/2} =
O(\hbar^{1-d} )$. 
On one hand, for $d>1$ and $\hbar \to 0$, $<(\Delta N_{0,n})^2>^{1/2}$
diverges.
On the other hand however, the number of states in any classical energy 
range which, in turn, corresponds to the size of the truncated Hamiltonian 
matrix, $N$, is $O(\hbar^{-d})$ and therefore diverges much faster. In fact,
${\nobreak <(\Delta N_{0,n})^2>^{1/2}/N = O(\hbar)}$ and accordingly, for small
enough $\hbar$, the $H_{nm}$ matrix is banded. Moreover, Eq. (\ref{eq:diag})
implies that the diagonal matrix elements, $H_{nn}$, grow on average
as the volume of the energy shell grows. Both these features are 
absent in the traditional Random Matrix Ensembles, e.g. GOE,
and this lack of
structure leads to extended eigenvectors. A Random Matrix model which does
include a simplified version of this structure is the Wigner
ensemble\cite{wig,f1920} which
is composed of banded matrices of band width, $b$, with diagonal matrix
elements that increase on average with constant rate, $\alpha$. Specifically,
$<h_{nm}>_e  = \alpha n \delta_{nm} $, where $<...>_e$ denotes
ensemble averaging, and $\sigma_{nm}^2 \equiv <h_{nm}^2>_e -<h_{nm}>_e^2 = 1
+\delta_{nm}$ for $|n-m| <b$ and vanishes otherwise. As in the case of
the GOE, it is assumed that the strongly chaotic nature of the classical
dynamics is sufficient to ensure that the matrix elements are uncorrelated
random variables. In what follows we shall use the Wigner ensemble and its
properties in order to understand the behavior of the eigenvectors of
$H_{nm}$.

For $\alpha=0$ the Wigner ensemble is equivalent to the Banded
Random Matrix Ensemble (BRME) which in turn can be thought of as 
an Anderson model with random, long-range hopping. Similarly, 
it is useful to interpret the Wigner ensemble at finite $\alpha$ as
an Anderson model under the influence of a constant electric field
of strength $\alpha$.
This approach enables one to visualize the behavior of the local
density of states which is closely related to that of the eigenvectors.
In the absence of electric field, the Anderson model is on average
translational invariant. Therefore, the ensemble averaged local 
density of states,
\begin{equation}
\rho_L (E,n) \equiv < \sum_i  |(u_i)_n |^2 \delta (E -E_i) >_e \ ,
\label{eq:rloc} \end{equation}
where $(u_i)_n$ is the $n$-th component of the $i$-th eigenvector
of the random matrix and $E_i$ the corresponding eigenvalue,
is proportional to the average density of states itself,
$\rho(E)$. In particular, for the 
BRME, both $\rho(E)$ and $\rho_L(E,n)$ are in the form of a 
semicircle of radius $2\sqrt{2b}$. While turning on the electric field
breaks the translational invariance, for small enough fields the
hopping  potential varies much faster than the electric one and the
adiabatic  approximation that relays on the separation of these two energy
scales is known as the {\it sloping band picture}. In this regime, one has 
at each site the same $\rho_L(E,n)$ as at zero field only that its 
center is shifted to include the additional electric energy that 
is increasing linearly along the lattice. This is schematically 
illustrated in Fig. \ref{fig:1} where the allowed region inside
the band represents the 
energy shell of the Wigner ensemble. While taking a section through
the energy band at a fixed position, $n$, gives $\rho_L(E,n)$, a 
section at a fixed energy is expected to give information on the 
behavior of the average eigenvector. In particular, one expects that
the average eigenvector is constrained to lie within the energy shell.
Independently of the sloping band picture, the $\rho_L(E,n)$ for the Wigner
ensemble was derived in Ref. \ref{r:wig} (see also Ref. \ref{r:ldos}).
It was found that
\begin{equation}
\rho_L (E,n)  = {1\over {\alpha b}} f ({{E -\alpha n}\over{\alpha b}}, q)
\ , \label{eq:rloc1} \end{equation}
where $q \equiv (\alpha^{2} b)^{-1}$.
For small electric fields, $q \gg 1$, the semicircle behavior persists
\begin{equation}
f( x, q) = (4\pi q)^{-1} \sqrt{8 q -x^2} \ , \label{eq:ql}
\end{equation}
and for large fields, $q \ll 1$, the profile is Lorentzian,
\begin{equation}
f( x, q) = { q \over {\pi^2 q^2 + x^2 }} \ . \label{eq:qs}
\end{equation}
In fact, Eqs. (\ref{eq:ql}-\ref{eq:qs}) only hold 
for $x < 1$. For $x \gg 1$, $f$ is the solution of an integral equation
for which 
\begin{equation}
f( x, q) \simeq c \exp [-2 x \ln (x e^{-1} \sqrt{2 q^{-1}
\ln (x/\sqrt{q}) } ) ] \ , \label{eq:tl}
\end{equation}
represents an approximate solution. 

We now turn to discuss the behavior of the average eigenvector,
that is defined as the average variance of the vector component at 
a fixed distance from the largest component, $g(l) \equiv <|(u_i)(n
- n_{max}))|^2>_e$. We find that the tails of $g(l)$ far
outside the energy shell are directly determined by the local density of
states, $\rho_L (E,n)$ (see Fig. \ref{fig:2}). On the other hand, inside the
energy shell the shape of $g(l)$ is different in the large and
small $q$ regimes. At large electric field, $q \ll 1$, the disorder
is too weak to localize the eigenvector and accordingly, like in the 
case of the tails, $g(l)$ is determined by the local density of states.
That is, it takes the Lorentzian form of Eq. (\ref{eq:ql}) with
$l = E/\alpha$. For weak field however, $q \gg 1$, the band is only
slightly sloped and as a consequence, the variation of the local 
density of states is slower than the scale on which localization
due to disorder takes place. Thus, the disorder is dominant and
the same exponentially localized shape as in the absence of the
electric field is obtained (see Fig. \ref{fig:3}). The transition
between the two regimes is centered
at $q_c$ where the hopping range is of the same size as the
energy shell width. A rough estimate of $q_c$ can be obtained
assuming that the width of the energy shell does not change with
$\alpha$ staying $4\sqrt{2b}$ all the way down to $q_c$. While such
estimate gives $q_c \approx 0.125$, numerically a value of $q_c
\simeq 0.09$ is obtained.\cite{cas2}

In order to establish the correspondence between the various
regimes of the Wigner ensemble and the structure of Hamiltonian
matrices, one needs to use the semiclassical formulas for the
parameters of the ensemble. Notice that the variances of the 
matrix elements in the Wigner ensemble are $O(1)$ unlike those in
$H_{nm}$ and therefore, the effective value of the electric field 
for the Hamiltonian matrix is $\alpha_{ef} = \alpha/\sigma_O $. 
Then, using Eqs. (\ref{eq:diag} - \ref{eq:dels}), one obtains that
$q = O(\hbar^{-2})$ implying that in the semiclassical limit
the $H_{nm}$ matrices are in the disorder dominated regime and
therefore exponentially localized eigenvectors can be found
whenever the system is sufficiently disordered. This is the main
result of the paper.

Let us now study the extent to which the Wigner ensemble is a
good model for a particular Hamiltonian matrix,\cite{fla} namely, that of
the Coupled Quartic Oscillators model,
\begin{equation}
H = {{ p_1^2 + p_2^2 }\over 2} + b_c q_1^4 + b_c^{-1} q_2^4
-a_k q_1^2 q_2^2  \ ,
\end{equation}
at $b_c = \pi/4 $ and $a_k =1.6$ where the classical dynamics
in the Poincare section appears to be fully chaotic to a
resolution of about $0.4\%$ of $\hbar$ which, in turn, was taken
to be unity. Moreover, we take $H_0$ to be composed of two 
uncoupled harmonic oscillators with frequencies $w_1 =4.11$ and
$w_2 = 1.3$, truncate the resulting $H_{nm}$ matrix to the 
first $N$ basis states, $N=800$,
and average the eigenvectors to obtain the corresponding $g(l)$
function (see Fig. \ref{fig:3}). The comparison with the prediction of the
Wigner ensemble that has the same average $b$ and $\alpha$ and the
same $N$ shows large quantitative disagreement. In particular, the
exponentially decaying shape quickly saturates into broad shoulders
that end in a fast drop corresponding to the edge of the energy shell. 
However, the reasons for this discrepancy are not related to the
intrinsic differences between evolution type matrices and Hamiltonian ones,
namely the behavior of the diagonal matrix elements since these are
irrelevant in the large $q$ regime. Instead, it is a consequence of the
fact that the $U_{nm}$ matrix of kicked systems is a lot more random,
and thus a lot closer to the BRME, than is the $H_{nm}$ matrix of our
example to the Wigner ensemble. In the case of the Kicked Rotor for example, 
\begin{equation}
U_{nm} = \exp (-i\hbar T n^2 /2) (-i)^{m-n} J_{m-n}(k) \ ,
\label{eq:kr}
\end{equation}
where $J$ are the Bessel functions. The exponential factor originates from the
kinetic energy term playing here the role of $H_0$ and it constitutes an
efficient source of randomness. In the $H_{nm}$ matrix on the other 
hand, the $H_0$ term is both additive and not exponentiated and
therefore, the mechanism for generating randomness in Eq. (\ref{eq:kr})
is absent here. It is natural to expect that, in order to fully recover
the predictions of the Wigner ensemble, one needs to use more complex bases
than that of harmonic oscillators as an alternative source of randomness
in the matrix elements of $H_{nm}$.\cite{var}

In summary, although eigenstates of time-independent Hamiltonian
systems are indeed localized inside the energy shell, the quantitative
behavior is controlled by the degree of correlation between matrix
elements. The study of the way in which such correlations are 
influenced by the choice of basis is an exciting open question to
be addressed in future work. In contrast to the case of the eigenvectors,
the eigenvalue statistics is greatly influenced by the behavior
of the diagonal matrix elements. Specifically, a finite $\alpha$ makes
neighboring eigenvalues correspond to eigenvectors that strongly
overlap and the ensuing level repulsion leads to a Wigner type 
spacing distribution.\cite{f1920}

We would like to thank B. Horovitz, D.M. Leitner and B. Shapiro for useful
discussions. This work was supported by the Israel Science Foundation
administered by the Israel Academy of Sciences and Humanities.

\newpage
\setcounter{totalnumber}{3}

\begin{figure}[h]
\bigskip
\caption{
\baselineskip 24pt
\hsize 15 truecm
\hoffset -1.0 truecm
The sloping band picture. The full lines correspond to the energy band
edges. The dashed lines indicate fixed $n$ sections which give the
local density of states and fixed energy sections which for small $q$
lead to the average eigenvector. }
\label{fig:1}
\end{figure}

\begin{figure}[h]
\caption{
\baselineskip 24pt
\hsize 15 truecm
\hoffset -1.0 truecm
The numerically obtained $g(l)$ function for the Wigner ensemble
with $\alpha= 2$ and $b=14$ ($q=0.0179$) ($\diamond$) are
compared with Eqs. (9 - 10). Moreover, the $\times$ symbols
represent the numerical ${1\over 4} g(l/4)$ function for the Wigner
ensemble with $\alpha= 1$ and $b=56$ (same $q$), verifying the
scaling of Eq. (7) in the $n$ direction. }
\label{fig:2}
\end{figure}

\begin{figure}[h]
\caption{
\baselineskip 24pt
\hsize 15 truecm
\hoffset -1.0 truecm
The average eigenvector for the Coupled Quartic Oscillators (full) when
the matrix has effective $\alpha_{ef}= 0.013$ and $b=12.9$ ($q = 446$) and
is compared with the corresponding result from the Wigner ensemble with
the same $\alpha$ and $b=13$ (dashed). The $\times$ symbols correspond to
the $g(l)$ of the Wigner ensemble with the same $b$ but $\alpha=0$,
indicating that in this regime, $\alpha$ has almost no influence on the
shape of the average eigenvector except for the sharp drop
at $l \approx \pm 780$ due to the band edge. }
\label{fig:3}
\end{figure}


\newpage

\begin{references}
\bibitem{fish}
S. Fishman, D.R. Grempel, and R.E. Prange, Phys. Rev. Lett. {\bf 49},
509 (1982).
\bibitem{f4}
M. Feingold, S. Fishman, D.R. Grempel, and R.E. Prange, Phys. Rev.
{\bf B31}, 6852 (1985).
\bibitem{bohi}
O. Bohigas and M.J. Giannoni, in {\it Mathematical and
Computational Methods in Nuclear Physics}, edited by J.S. Dehesa,
J.M.G. Gomez, and A. Polls, Lecture Notes in Physics
Vol. 209 (Springer-Verlag, Berlin, 1984).
\bibitem{cas1}
G. Casati, L. Molinari, and F. Izrailev, Phys. Rev. Lett.
{\bf 64}, 1851 (1990).
\bibitem{ber1}
M.V. Berry, in {\it Chaotic Behavior of Deterministic Systems },
Vol. 36 of {\it Les Houches Summer School Proceedings}, edited by G. Iooss,
R.H.G. Helleman, and R. Stora (North-Holland, Amsterdam, 1983).
\bibitem{diag}
A.I. Shnirelman, Usp. Mat. Nauk. {\bf 29}/6, 181 (1974); M. Feingold,
N. Moiseyev, and A. Peres, Chem. Phys. Lett. {\bf 177}, 344 (1985).
\bibitem{f8}
M. Feingold and A. Peres, Phys. Rev. {\bf A34}, 591 (1986).
\bibitem{wig}
E.P. Wigner, Ann. Math. {\bf 62}, 548 (1955); {\bf 65} 203 (1957).
\label{r:wig}
\bibitem{f1920}
M. Feingold, D.M. Leitner, and M. Wilkinson, Phys. Rev. Lett.
{\bf 66}, 986 (1991); J. Phys. A {\bf 24}, 175 (1991).
\bibitem{ldos}
D.M. Leitner and M. Feingold, J. Phys. A {\bf 26}, 7367 (1993);
G. Casati, B.V. Chirikov, I. Guarneri, and F.M. Izrailev, Phys.
Rev. E {\bf 48}, 1613 (1993); Y.Y Fyodorov, O.A. Chubykalo, F.M. Izrailev,
and G. Casati, Phys. Rev. Lett. {\bf 76}, 1603 (1996).
\label{r:ldos}
\bibitem{cas2}
A similar picture was independently suggested by G. Casati, B.V. Chirikov,
I. Guarneri, and F.M. Izrailev, preprint (1995).
\bibitem{fla}
A previous such comparison was done using a model of the Ce atom in
V.V. Flambaum, A.A. Gribakina, G.F. Gribakin, and M.G. Kozlov,
Phys. Rev. A{\bf 50}, 267 (1994). Their model has $12$ degrees of freedom
and no clear classical limit. While, the value of $q$ which they find
is close to $q_c$, $q \simeq 0.18$, some agreement with the predictions
of the corresponding Wigner ensemble is observed.
\bibitem{var}
Another difference between the Wigner ensemble and the $H_{nm}$ matrix 
studied here is that the latter is not semiclassical enough leading to
nonnegligible variation of the band width and mean spacing with $E_0$.

\end{references}
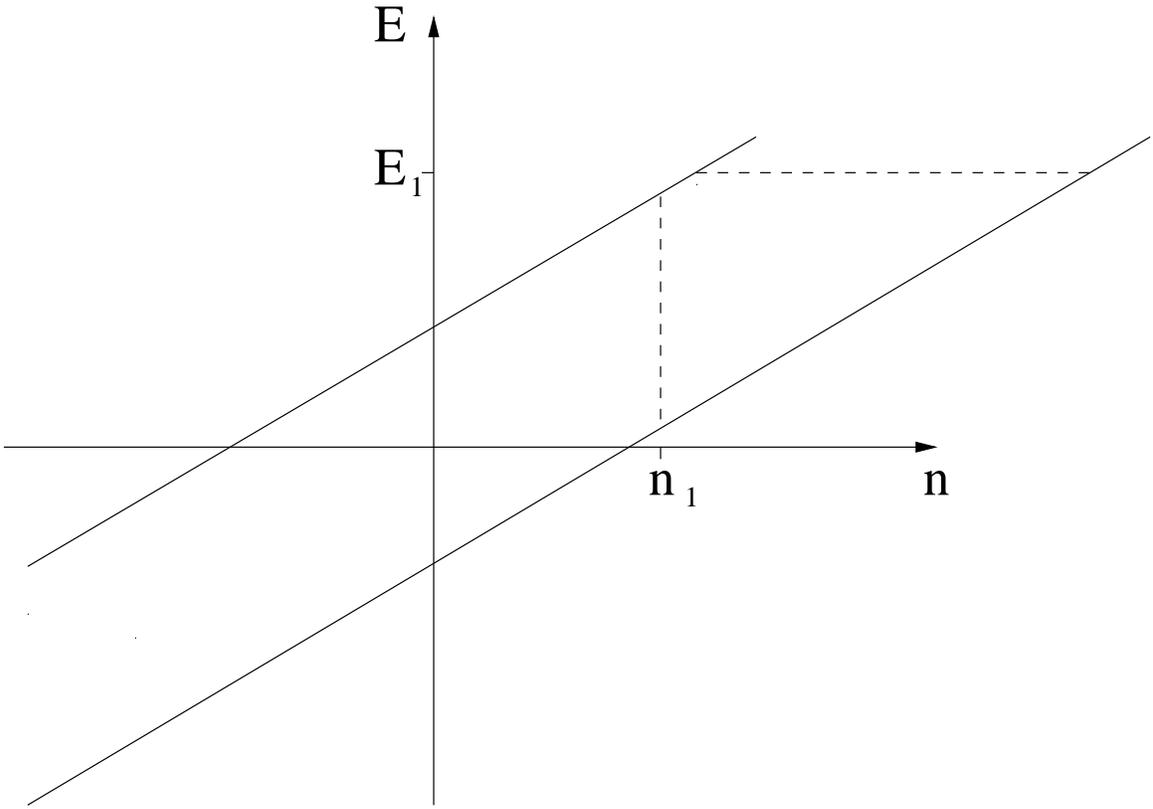
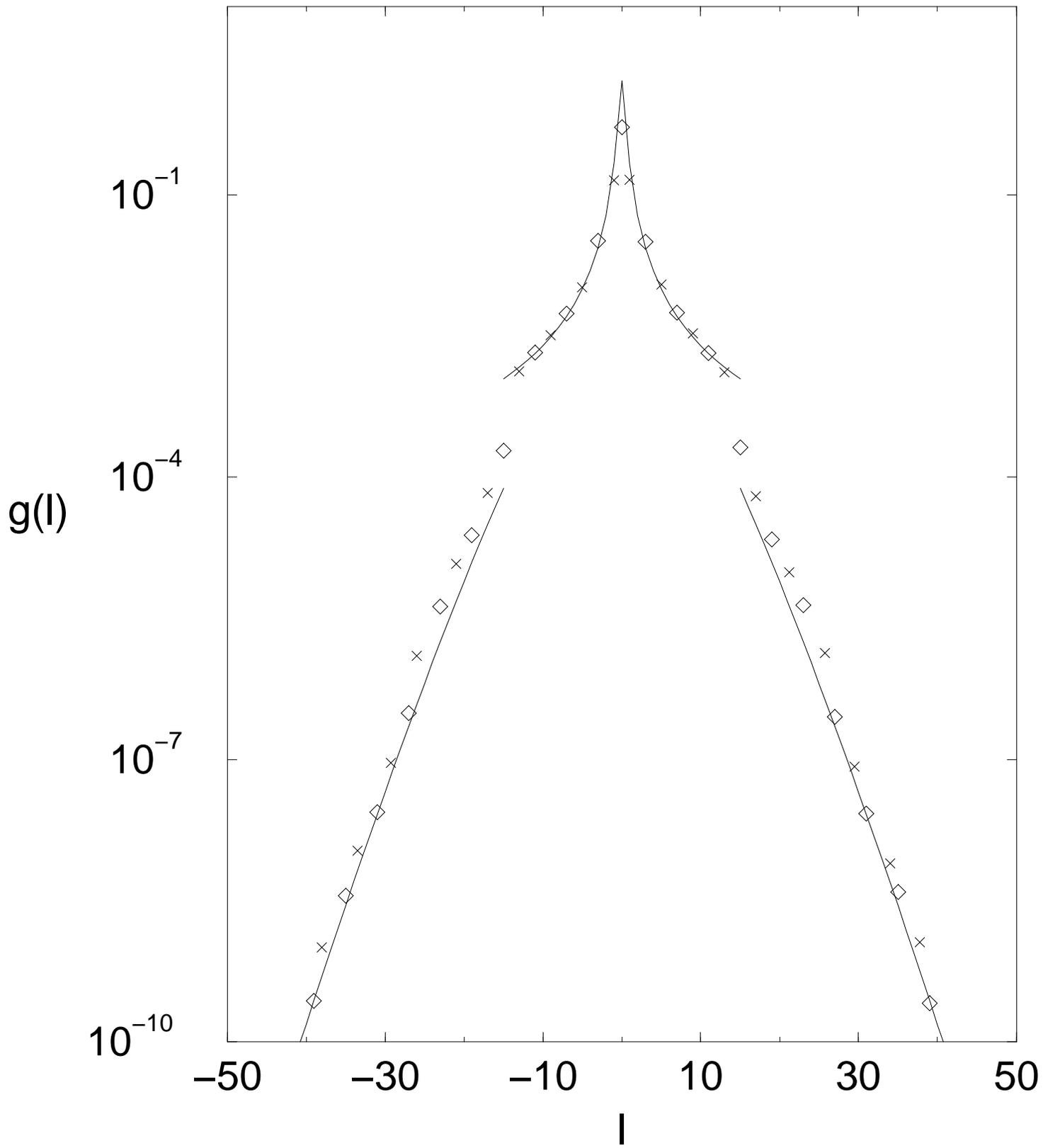
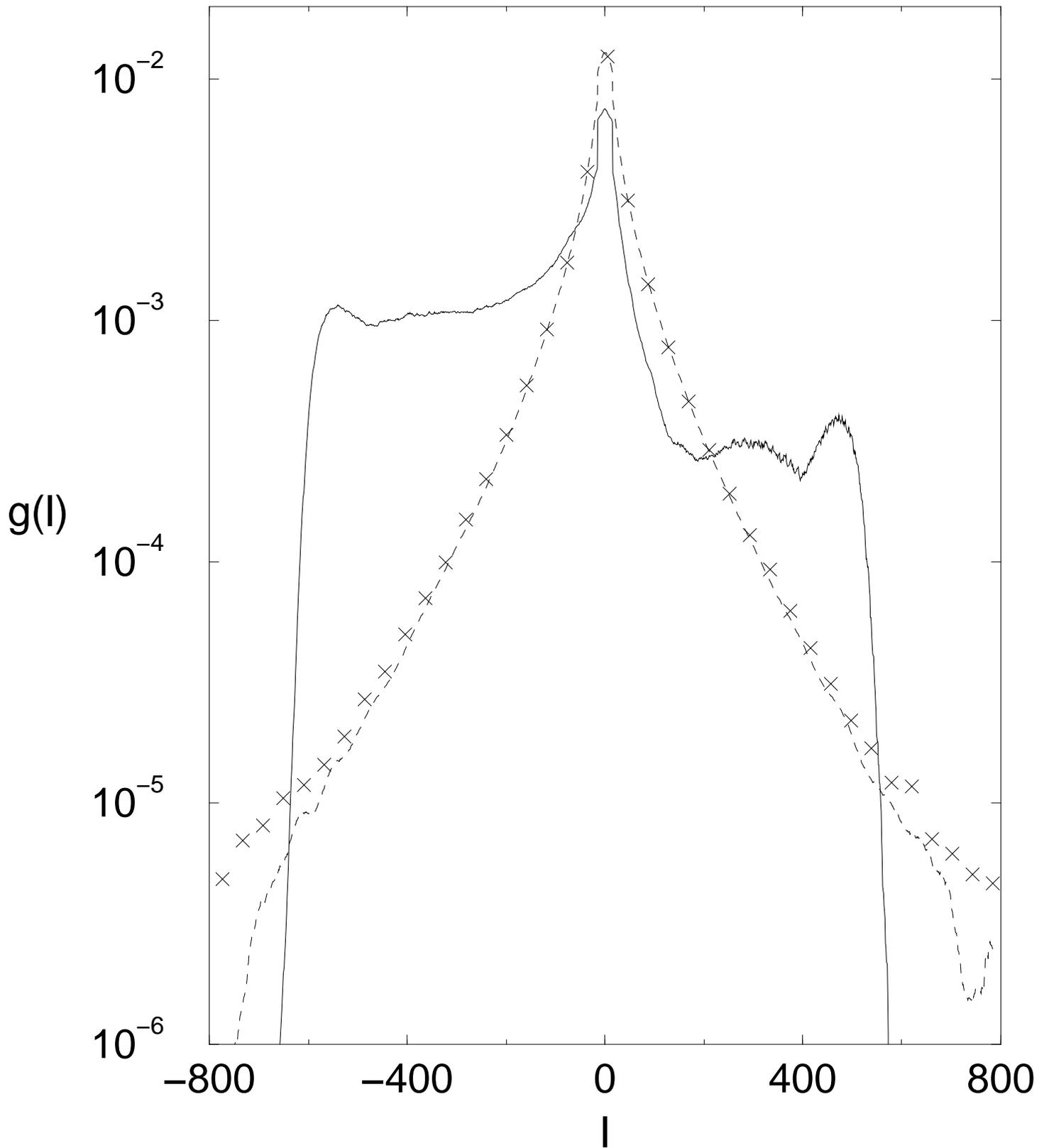
\end{document}